\documentclass[aps,amsmath,amssymb,prb,showpacs,showkeys,superscriptaddress,reprint,floatfix]{revtex4-1}
\usepackage[USenglish]{babel}
\usepackage{amssymb}
\usepackage{amsmath}
\usepackage{graphicx}
\usepackage{color}
\usepackage{multirow}
\usepackage[toc,page]{appendix}
\usepackage{natbib}

\begin{document}
\title{The effect of hydrostatic pressure and uniaxial strain on the electronic structure of Pb$_{\text{1-x}}$Sn$_{\text{x}}$Te}

\author{Matthias~Geilhufe}
\affiliation{Max Planck Institute of Microstructure Physics, Weinberg 2, 06120 Halle, Germany}

\author{Sanjeev~K.~Nayak}
\email{sanjeev.nayak@physik.uni-halle.de}
\affiliation{Institute of Physics, Martin Luther University Halle-Wittenberg, Von-Seckendorff-Platz 1, 06120 Halle, Germany}

\author{Stefan~Thomas}
\affiliation{Max Planck Institute of Microstructure Physics, Weinberg 2, 06120 Halle, Germany}

\author{Markus~D\"ane}
\affiliation{Physical and Life Sciences, Lawrence Livermore National Laboratory, PO Box 808, L-372, Livermore, CA 94551, United States}

\author{Gouri~S.~Tripathi} 
\affiliation{Physics Department, Berhampur University, Berhampur 760007, Odisha, India}

\author{Peter~Entel}
\affiliation{Faculty of Physics and CENIDE, University of Duisburg-Essen, 47048 Duisburg, Germany} 

\author{Wolfram~Hergert}
\affiliation{Institute of Physics, Martin Luther University Halle-Wittenberg, Von-Seckendorff-Platz 1, 06120 Halle, Germany}

\author{Arthur Ernst}
\affiliation{Max Planck Institute of Microstructure Physics, Weinberg 2, 06120 Halle, Germany}


\begin{abstract}
The electronic structure of Pb$_{1-x}$Sn$_{x}$Te is studied by using the relativistic Korringa-Kohn-Rostoker Green function method in the framework of density functional theory. For all concentrations $x$, Pb$_{1-x}$Sn$_{x}$Te is a direct semiconductor with a narrow band gap. In contrast to pure lead telluride, tin telluride shows an inverted band characteristic close to the Fermi energy. It will be shown that this particular property can be tuned, first, by alloying PbTe and SnTe and, second, by applying hydrostatic pressure or uniaxial strain. Furthermore, the magnitude of
strain needed to switch between the regular and inverted band gap can be tuned by the alloy composition. Thus, there is range of potential usage of Pb$_{1-x}$Sn$_{x}$Te for spintronic applications.
\end{abstract}
\maketitle

\section{Introduction}

Among the narrow band gap semiconductors, SnTe and PbTe are in focus since long. The interest has been driven by their applications in electronic industry, essentially because of the effective control of charge carriers achieved in these systems. Tin and lead telluride alloys are attractive due to their use as infrared photodetectors~\cite{Khokhlov-2000,Rogalski-2003}. Furthermore, PbTe and SnTe play an important role in the fabrication of thermoelectric materials~\cite{He-2015}. Especially by alloying, e.g. with bismuth (Bi) or antimony (Sb) a high figure of merit can be reached \cite{snyder-2008, Dmitriev-2010}. Recently, there has been revived interest in the system due to carrier controlled ferromagnetism observed by suitable doping~\cite{Story-1996, Nadolny-2002, Jonge-1991}. SnTe and its alloy with PbTe is found to be a topological crystalline insulator, where robust conducting states are observed in their electronic structures~\cite{Hsieh-2012,Tanaka-2012}. A topological surface state is also predicted for PbTe--SnTe multilayers~\cite{Yang-2014}.

The topological characteristic in the electronic structure is related to a fundamental band inversion at the Fermi level observed for tellurides like HgTe and SnTe. In the pure solids, both PbTe and SnTe have a direct band gap at the $L$-point of the Brillouin zone. However, whereas the analysis of the relativistic wave functions for PbTe show an even symmetry ($L_{6}^{+}$) at the valence band edge and an odd symmetry ($L_{6}^{-}$) at the conduction band edge, the associated irreducible representations of the states at the band edges for SnTe are arranged in reverse manner~\cite{Bernick-1970}. As a result, the band gap of PbTe is said to be positive, whereas the band gap of SnTe is treated as negative.

There exist various experimental reports on the change of the band gap with concentration $x$ in Pb$_{1-x}$Sn$_x$Te\cite{Dimmock-1966,Ferreira-1999}. It is found that by increasing the concentration of Sn the band gap decreases linearly up to a certain concentration $x_0$ where the band gap vanishes. By further increasing $x$ the band gap starts to increase linearly until the band gap of pure SnTe is obtained. In the past, the physical properties of the alloy have been studied theoretically by means of the $\vec{k}\cdot\vec{\pi}$ perturbation method, where the band gap is treated as a function of the concentration $x$ and the temperature~\cite{gstpbteandalloy,gstalloy}. Alloy properties were also studied by means of the tight-binding method~\cite{Safaei-2013,Rauch-2013}. An \textit{ab initio} investigation of Pb$_{1-x}$Sn$_x$Te was reported by Gao and Daw~\cite{Gao-2008}, where a quasirandom distribution of components within a supercell was used to study the alloy~\cite{Zunger-1990,Wei-1990}. 

A conventional supercell approach is restricted by the number of atoms and only allows for particular concentrations. To study the electronic properties of the Pb$_{\text{1-x}}$Sn$_{\text{x}}$Te alloy within the whole concentration range $0 \ge x \ge 1$ the coherent potential approximation (CPA) was applied which has been used widely for the investigation of chemical and spin disordered alloys~\cite{Temmerman-1978,Faulkner-1980,sprkkr1}. The main focus of this study is to investigate the size as well as the character of the band gap in dependence of the application of hydrostatic pressure and uniaxial strain. In contrast to hydrostatic pressure, the application of uniaxial strain is technologically more relevant in thin-film and multi-layers samples.

The outline of the paper is arranged as follows: After introducing the computational methods in section~\ref{section:methods} we discuss the electronic structure of pure PbTe and SnTe in section~\ref{ElectronicStructureSnTe-PbTe}. The electronic structure of the Pb$_{1-x}$Sn$_x$Te alloy as well as the influence of hydrostatic pressure is studied in section~\ref{secalloy}. Finally, the influence of uniaxial strain is discussed in section~\ref{secstrain}. 

\section{Computational methods\label{section:methods}}
The electronic structure calculations were performed by means of the Korringa-Kohn-Rostoker Green function method (KKR)~\cite{sprkkr1} in the framework of density functional theory~\cite{Hohenberg-1964, Kohn-1965}, as implemented within the computer code \textit{Hutsepot}~\cite{Ernst2007,Luders-2001}. The atomic potentials of the scattering centres were obtained self-consistently by applying the scalar relativistic approximation~\cite{Koelling-1977,Takeda-1978} and the atomic sphere approximation~\cite{Temmerman-1978}. The evaluation of the Green function in terms of spherical harmonics was expanded up to a maximal angular momentum of $l=3$. The energy contour along the complex energy plane consisted of 24 points. As an approximation for the exchange-correlation functional we used the LibXC\cite{marques-2012} implementation of PBE~\cite{pbe}, which is a generalized gradient approximation. To properly simulate the Pb$_{1-x}$Sn$_x$Te alloy the coherent-potential approximation~\cite{Soven-1967,Velicky-1968,Gyorffy-1972} was applied. Band structures were calculated in terms of the Bloch spectral function by using a fully-relativistic KKR method based on the solution of the Dirac equation. The relativistic KKR code was developed in order to describe the relativistic effects like spin-orbit coupling in a suitable manner. Details can be found in Ref.~\onlinecite{Geilhufe-2015}.

Pb$_{1-x}$Sn$_x$Te crystallizes in rock salt crystal structure. For the investigation of hydrostatic pressure on the electronic structure a face-centred cubic (fcc) primitive cell is chosen, consisting of one Te atom as well as one Pb or Sn atom, respectively. To improve convergence properties of the Green function, two additional empty spheres were added at non-occupied crystallographic sites. To simulate uniaxial strain we chose a simple cubic (sc) unit cell, which contains four times as many atoms as the primitive fcc cell. The lattice vectors are given by $\vec{a}=(a,0,0)$, $\vec{b}=(0,b,0)$ and $\vec{c}=(0,0,c)$. The relation $a_0=b_0=c_0$ holds for the equilibrium lattice constants. The associated dimensions of the $\vec{k}$-mesh within the Brillouin zone were $20 \times 20 \times 20$ for the fcc cell, as well as $12 \times 12 \times 12$ for the sc cell.

The atomic sphere approximation is a coarse approximation for estimating equilibrium lattice constants. We thus employed the Vienna \textit{ab initio} simulation package (VASP) for estimating the lattice parameters of the pure systems PbTe and SnTe. For the calculations $8\times8\times8$ $\vec{k}$-point mesh was chosen. Calculations were done using the high performance mode using a plane wave energy cut-off of 350~eV. The lattice parameters of the alloy  are calculated from Vegard's law \cite{vegard-1921} which states that the lattice constant changes linearly for $0 < x <1$. Test calculations with VASP using a large unit cell agreed to the approximation fairly well.

\section{Results and discussion\label{section:result_and_discussion}}

\subsection{Electronic structure of PbTe and SnTe}\label{ElectronicStructureSnTe-PbTe}

The lattice parameters and the band gap for PbTe and SnTe obtained with VASP are tabulated in Tab.~\ref{tab-1}. The equilibrium lattice constants are found to be slightly larger than those reported in experiments~\cite{Mariano-1967,Dalven-1969}. It was observed that spin-orbit interaction (SOI) does not influence the lattice constants. The band gap of PbTe and SnTe are estimated to be 0.81~eV and 0.05~eV in GGA. With the incorporation of SOI the valence bands and conduction bands split. The splitted bands effectively move towards each other thereby reducing the band gap to $0.07$~eV and $-0.11$~eV for PbTe and SnTe, respectively (refer Tab.~\ref{tab-1}). Here, the negative band gap signifies the inversion of symmetry of the states at the band edges. Although the splitting for SnTe is smaller than for PbTe, the smaller non-relativistic band gap of SnTe allows band inversion. In agreement with previous theoretical studies~\cite{Tung-1969,Tung-1970,Rabii-1969} it was found that SOI is crucial for reducing the band gap for both compounds with characteristic band inversion for SnTe.

 \begin{table}[b!]
  \def\arraystretch{1.2}
  \caption{Equilibrium lattice parameters and band gap calculated using VASP in comparison with experimental values. Calculations were performed with and without spin-orbit interaction, denoted as GGA and GGA+SOI, respectively.\label{tab-1}}
  \begin{ruledtabular}
   \begin{tabular}{l l l l l}
    \multirow{2}{*}{Method} & \multicolumn{2}{c}{PbTe} & \multicolumn{2}{c}{SnTe}  \\  
    \cline{2-3} \cline{4-5}
                              & $a$ ({\AA}) & $E_g$ (eV) & $a$ ({\AA}) & $E_g$ (eV) \\
\hline
GGA      &  $6.565$ & $0.81$ &    $6.408$         &  $0.05$   \\
GGA + SOI&  $6.562$ & $0.07$ &    $6.410$         &  $-0.11$  \\
Expt.    &  $6.462$~\cite{Mariano-1967} & $0.19$~\cite{Foley-1977} &  $6.303$~\cite{Dalven-1969}  &  $-0.2$--$-0.3$~\cite{Tsu-1968}  \\
         &             &  $0.32$~\cite{Scanlon-1959}  &    & \\
   \end{tabular}
  \end{ruledtabular}
 \end{table}

In the framework of the fully relativistic Green function method, the Bloch spectral function (BSF) $A_{\text{B}}(\vec{k},E)$ is calculated which is given by~\cite{weinberger-1990}
 \begin{multline} A_{\text{B}}(\vec{k},E) = \\-\frac{1}{\pi} \operatorname{Im} \operatorname{Tr} \left[\sum_{\vec{R}_j\in \mathcal{L}} e^{i \vec{k}\cdot \vec{R}_j}\int \mathrm{d}^3 r_j\, \underline{\underline{G}}(\vec{r}_j,\vec{r}_j+\vec{R}_j,E) \right].  \label{eqbsf} \end{multline} 
 The summation on the right-hand side of the equation extends over all vectors $\vec{R}_j$ of the lattice $\mathcal{L}$. The trace is a summation of the diagonal components of the $4\times 4$ dimensional multiple-scattering Green function~\cite{Huhne-1998} $\underline{\underline{G}}$ and $E$ denotes the energy.  \begin{figure}[t]
  \centering
\includegraphics[width=0.49\textwidth]{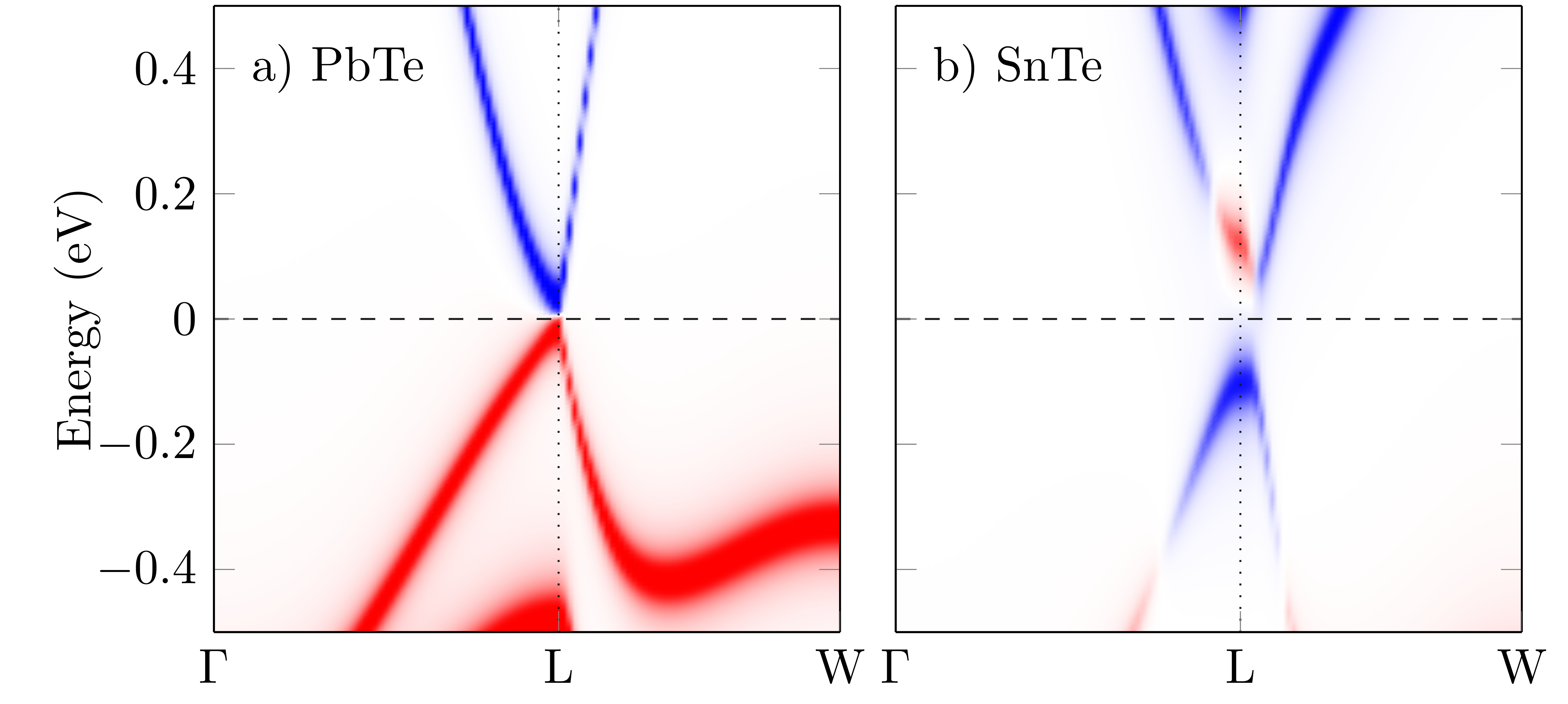}
 \caption{The difference between the cation and the anion contribution within the Bloch spectral function ($\Delta A_{\text{B}}$) plotted for the $\Gamma-L-W$ high-symmetric line in the Brillouin zone for PbTe and SnTe, respectively. The Fermi energy is set to zero. The red color signifies that the main contribution to the BSF is from the anion atom (Te), while the blue color signifies that the main contribution is from the cation atom (Pb or Sn). 
 \label{figbsfconc}}
\end{figure}

In the band structure of a regular semiconductor, the valence band is made up of anion character while the conduction band is made up of cation character. In band-inverted semiconductors this property is reversed. Thus, the difference of the site-resolved BSF from the cation and the anion, 
\begin{equation}
\Delta A_{\text{B}} = A_{\text{B}}^{\text{Cation}}-A_{\text{B}}^{\text{Anion}},
\end{equation}
can be used as a tool for analysis of band inversion properties. In Fig.~\ref{figbsfconc}~a) and Fig.~\ref{figbsfconc}~b), $\Delta A_{\text{B}}$ is plotted for PbTe and SnTe, respectively. From the intermixed colours near the Fermi level which is set to zero the band inversion characteristic of SnTe is clearly revealed.

\subsection{Electronic structure of Pb$_{1-x}$Sn$_{x}$Te and hydrostatic pressure\label{secalloy}}
In our studies of the alloy we stick to the rock salt crystal structure over the whole range of concentration, even though a few reports on a temperature induced rhombhohedral phase exist~\cite{Yusheng-1985A,Yusheng-1985B,Yusheng-1985C}. The electronic structure of the Pb$_{1-x}$Sn$_{x}$Te was investigated for various concentrations $x$ and various lattice constants. The band gap energy was estimated by calculating the BSF [Eq.~\eqref{eqbsf}] at the L-point of the Brillouin zone within a small energy region $[E_F-\Delta, E_F + \Delta]$ ($\Delta \approx 0.2$~eV) about the Fermi energy $E_F$ using $N_E = 700$ energy points. All together the calculations were performed for 11 different concentrations and 20 different lattice parameters each, to produce a reasonable array of data. Since it is not possible to resolve the band gap in the region where the band gap changes sign, this array was interpolated using \textit{Mathematica}~\cite{math} to obtain a well defined curve. 

\begin{figure}[b!]
 \includegraphics[scale=0.65]{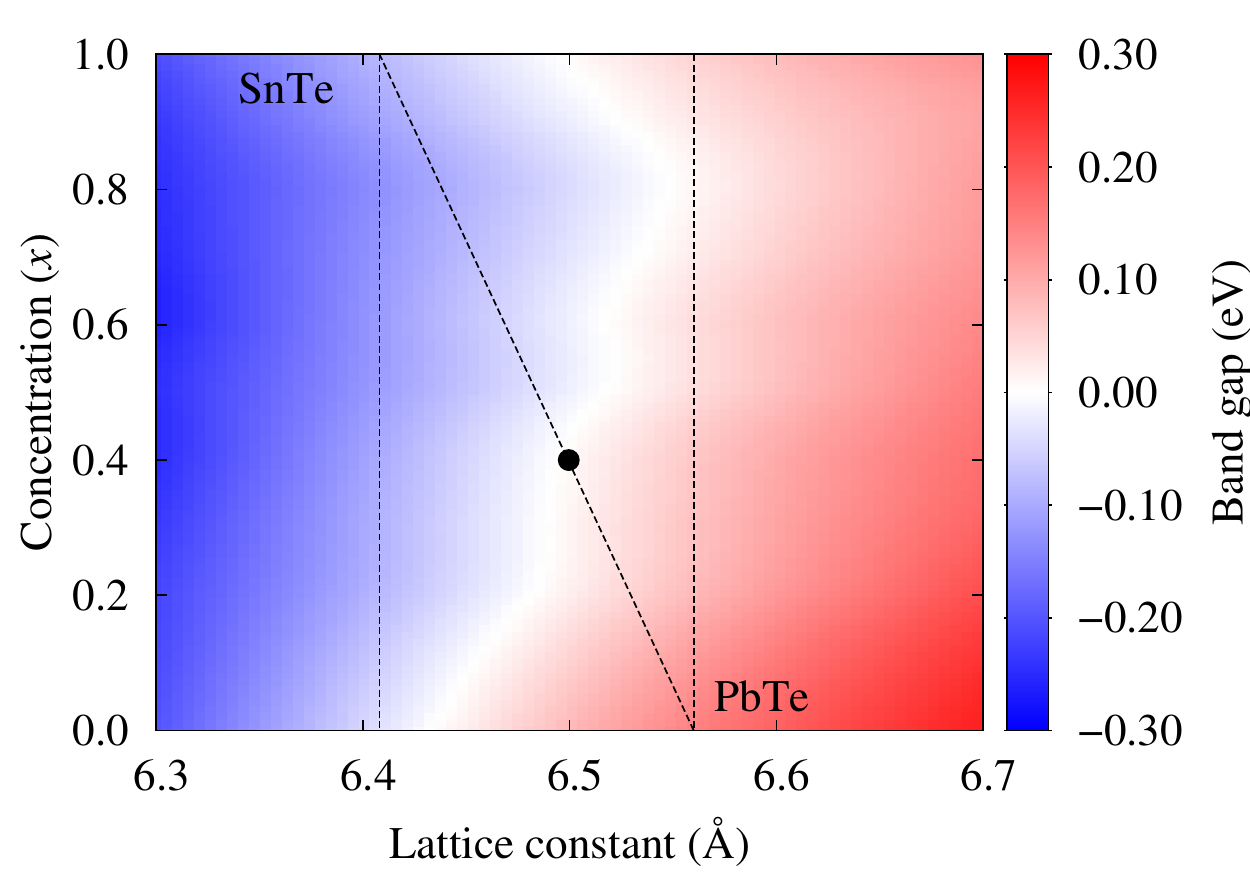}
 \caption{Band gap as a function of the lattice constant and the concentration $x$ in Pb$_{1-x}$Sn$_{x}$Te. The gradient shade with red color represents regular band gap and the blue color represents the inverted band gap. The dashed vertical lines represent the equilibrium lattice parameters of SnTe and PbTe and the straight dashed line across represents the lattice parameters versus concentration according to Vegard's law~\cite{vegard-1921}. The dot represents $E_g = 0$ along 
 the line obeying Vegard's law.\label{plot3d}}
\end{figure}

The result in terms of band gap versus concentration and lattice constant is illustrated in Fig.~\ref{plot3d}. The vertical dashed lines show the equilibrium lattice parameter of SnTe and PbTe. Following the line of Vegard's law (angular dashed line) from pure PbTe to pure SnTe, it can be seen, that with increasing amount of Sn the band gap energy decreases until it vanishes for a Sn concentration of $\approx 40 \%$ (vanishing intensity of red shade). This is in good agreement with experimental reports~\cite{Ishida-2010,Dixon-1968,Dimmock-1966}. Lowering the Pb concentration further increases the band gap, but with an inverted band characteristic at the L-point (growing intensity of blue shade). Hence, the size of the band gap as well as the band characteristic of Pb$_{1-x}$Sn$_{x}$Te can be tuned with changing the concentration of Pb.

Starting again with pure PbTe and leaving the concentration constant, a similar behaviour can be seen for compressing the crystal. Again, compression leads to a decrease of the band gap until the band gap vanishes at a lattice constant of about 6.43~{\AA}. This value corresponds to a lowering of the lattice constant of about 2\%. By using the Murnaghan equation of states, 
\begin{equation}
P(V) = \frac{B_0}{B'}\left[\left(\frac{V}{V_0}\right)^{-B'}-1\right],
\end{equation}
together with the values $B_0=37.73$~GPa and $B'=3.52$ obtained from our calculations, which are in good agreement with reported values for PbTe~\cite{Yang-2012}, it can be verified that these conditions can be obtained by applying a pressure of $P_c \approx 2.63$ GPa. Further compression will open the band gap again but with inverted band characteristic. Our estimation of the critical pressure $P_c$ is in very good agreement with the experimental value of 3~GPa~\cite{Ovsyannikov-2009}. According to Fig.~\ref{plot3d}, $P_c$ can be tuned by varying $x$ in the Pb$_{1-x}$Sn$_{x}$Te alloy. 

In order to understand the electronic effects for the band gap change for the Pb$_{1-x}$Sn$_{x}$Te alloy, we refer to the expression of the band gap achieved by means of the tight-binding model~\cite{Barone-2013},
\begin{equation}
E_g \approx \Delta_{0} + 10 t_{sp}^2 (\Delta_{1}^{-1}-\Delta_{2}^{-1})/3,\label{eqEg}
\end{equation}
where the abbreviations
\begin{align}
\Delta_{0} =&~ \bar{\epsilon}_{p, Pb/Sn} - \bar{\epsilon}_{p, Te}, \nonumber \\[3mm]
\Delta_{1}^{-1} =&~ \bar{\epsilon}_{p, Pb/Sn} - \epsilon_{s, Te}\quad\text{, and} \nonumber \\[3mm]
\Delta_{2}^{-1} =&~ \bar{\epsilon}_{p, Te} - \epsilon_{s, Pb/Sn}. \nonumber
\end{align}
are used. 
Here, $\epsilon_{l, X}$ and $\bar{\epsilon}_{l, X}$ are the orbital eigenvalues obtained by a non-relativistic and a relativistic treatment for the quantum number $l$ of element $X$. They are related by, $\bar{\epsilon}_{l, X} = \epsilon_{l, X}\pm 2\lambda_{X}$, where $\lambda_{X}$ is the spin-orbit coupling strength of element $X$.
In principle, there are two competing effects which influence the band gap, 
1) change in $\Delta_{0}$ and $t$ through the volume of the unit cell 
and 2) the change of the SOI-strength $\lambda$ through the concentration $x$. 
It has been suggested that the former criterion is decisive for the band gap change versus concentration~\cite{Barone-2013}. 

From the results we conclude that both effects cause a linear change of the band gap energy. In fact, a linear scaling of the band gap among the values $-0.11$~eV to $0.07$~eV for $x = 1$ and $x = 0$ gives zero band gap at $x = 0.39$, which is very close to the value estimated from KKR CPA calculations, $x = 0.4$. Our conclusion is in agreement with Wexler and Morgan~\cite{Wexler-1972} who studied  Pb$_{1-x}$Sn$_{x}$Te using a strain field and 
including relativistic effects in the framework of an augmented plane-wave method.
It must be noted that the SOI modeled separately also leads to an almost linear 
change of the lattice constant~\cite{Radzynski-2009}. In accordance to Ref.~\onlinecite{Barone-2013}, this emphasizes that the band inversion characteristics can be controlled by changing the volume. 

In summary, a linear behaviour for the change of the band gap can be observed along the line of Vergard's law, where the equilibrium lattice constant changes linearly with respect to the composition of the alloy. The same holds true 
for the change of the band gap for compressed cells below the equilibrium volume but parallel to Vegard's line. However, for larger volumes (parallel to Vegard's line), the SOI induces nonlinear effects which are reflected as the white region
in Fig.~\ref{plot3d}, where $E_g = 0$ appears nonlinearly.  

\subsection{Strained Pb$_{1-x}$Sn$_{x}$Te alloy\label{secstrain}}

\begin{figure}[t]
\includegraphics[scale=0.9]{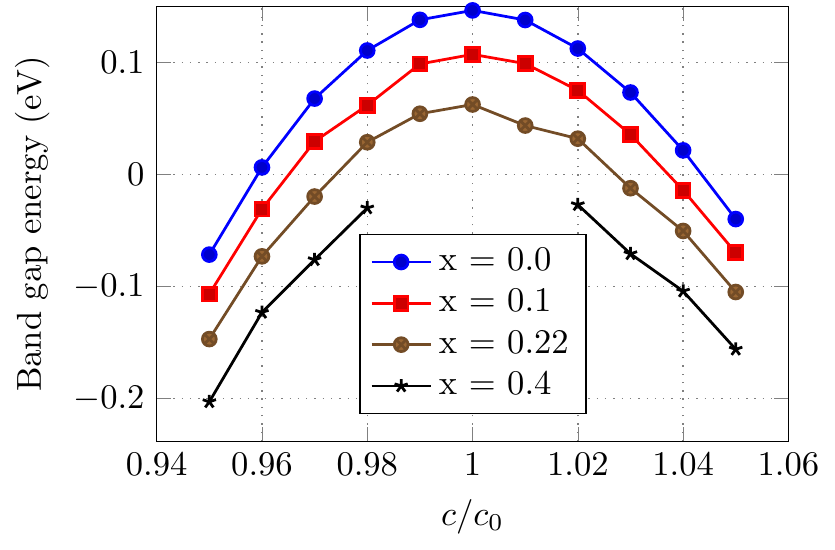}
 \caption{\label{plotstrain}  Band gap of Pb$_{1-x}$Sn$_{x}$Te as a function of $c/c_0$ obtained from fully-relativistic Green function calculations.} 
\end{figure}
Apart from the effect of hydrostatic pressure on solids, uniaxial and biaxial strain in the alloy is technologically more relevant. Thus, we have undertaken the study of band inversion characteristics in Pb$_{1-x}$Sn$_{x}$Te upon the action of uniaxial strain. Our attention is towards the physically interesting concentration range of $60$~\% Pb and higher (refer to Fig.~\ref{plot3d}). Therefore, the concentrations $100$~\%, $90$~\%, $78$~\% and $60$~\% were chosen to investigate how uniaxial strain influences the band gap. The unit cell was compressed or strained along the crystallographic $\vec{c}$-direction. The value of $a$ ($=b$) is thereby derived with the assumption that the volume of the unit cell is fixed as $c$ is changed. This approximation is reasonable, since the change of the volume is sufficiently small with respect to the concentration $x$.

The band gap as a function of $c$ for the above mentioned concentrations is shown in Fig.~\ref{plotstrain}. It can be verified that both, compression as well as strain leads to a lowering of the band gap with a parabolic dependence of $c/c_0$. In both cases, band inversion can be triggered if the change of the lattice constant $c$ is reasonably large. For pure PbTe the band gap vanishes for the values $c/c_0\approx 0.96$ and $c/c_0\approx 1.045$. By alloying, the band gap at the equilibrium changes $(a=b=c=c_0)$ and therefore the transition between inverted and non-inverted band characteristic already occurs for smaller deformations. For a Pb concentration of $60$\%, the unstrained Pb$_{x}$Sn$_{1-x}$Te is metallic. Our results show that by applying tensile or uniaxial strain it is possible to induce an inverted band gap into the alloy.

This conclusion is in agreement with the study of Qian~\textit{et al.} who did a similar investigation for bulk SnTe~\cite{Qian-2015}. Furthermore, they proposed that strained SnTe membranes can be used as frequency resolved infrared photodetectors. By choosing samples of Pb$_{x}$Sn$_{1-x}$Te and gradually changing the composition a comparable device can be constructed which is less fragile in comparison to thin SnTe membranes.
 
Two physical parameters, namely, strain and alloy composition allow tuning the band gap and the spin-texture of band edge states. Potential applications of this alloy lie in switching between a topological and a regular semiconductor by applying strain in spintronic devices. Furthermore, 
the effective strain can be controlled by choosing a proper composition of the alloy. Thus, besides the application as infrared photodetectors for a wide range of frequencies the tuning of band inversion leads to exciting possibilities
for novel functional devices such as spintronic-switches.
\section{Conclusions}\label{secconclusion}

We have studied the band inversion characteristics of Pb$_{1-x}$Sn$_{x}$Te using the relativistic KKR method combined with the CPA. In agreement with experiments it is found that an inversion of the band characteristic at the $L$-point occurs for a lead concentration of 60~\%. By applying hydrostatic pressure or uniaxial strain, the band gap can be tuned as well. Whereas the change of the band gap is more or less linear close to the equilibrium lattice constants for hydrostatic pressure, the change of the band gap shows a parabolic behaviour for the application of uniaxial strain. Hence, for the latter, band inversion can be obtained by either compressing or straining the sample along the crystallographic $\vec{c}$-direction.

As a consequence, the application of strain or hydrostatic pressure to Pb$_{1-x}$Sn$_{x}$Te allows to switch between a topological and a regular semiconductor. In this connection, the magnitude of the needed external pressure can be controlled by changing the composition of the alloy. Therefore, Pb$_{1-x}$Sn$_{x}$Te is a prominent candidate for the fabrication of topological-switches.

\section*{Acknowledgements}

The financial  support  from  Deutsche  Forschungsgemeinschaft through the framework of SFB762 ``Functionality of Oxide Interfaces'' and through the priority program SPP 1666 ``Topological Insulators'' is acknowledged. Lawrence Livermore National Laboratory is operated by Lawrence Livermore National Security, LLC, for the U.S. Department of Energy, National Nuclear Security Administration under Contract DE-AC52-07NA27344.


\begin{thebibliography}{59}%
\makeatletter
\providecommand \@ifxundefined [1]{%
 \@ifx{#1\undefined}
}%
\providecommand \@ifnum [1]{%
 \ifnum #1\expandafter \@firstoftwo
 \else \expandafter \@secondoftwo
 \fi
}%
\providecommand \@ifx [1]{%
 \ifx #1\expandafter \@firstoftwo
 \else \expandafter \@secondoftwo
 \fi
}%
\providecommand \natexlab [1]{#1}%
\providecommand \enquote  [1]{``#1''}%
\providecommand \bibnamefont  [1]{#1}%
\providecommand \bibfnamefont [1]{#1}%
\providecommand \citenamefont [1]{#1}%
\providecommand \href@noop [0]{\@secondoftwo}%
\providecommand \href [0]{\begingroup \@sanitize@url \@href}%
\providecommand \@href[1]{\@@startlink{#1}\@@href}%
\providecommand \@@href[1]{\endgroup#1\@@endlink}%
\providecommand \@sanitize@url [0]{\catcode `\\12\catcode `\$12\catcode
  `\&12\catcode `\#12\catcode `\^12\catcode `\_12\catcode `\%12\relax}%
\providecommand \@@startlink[1]{}%
\providecommand \@@endlink[0]{}%
\providecommand \url  [0]{\begingroup\@sanitize@url \@url }%
\providecommand \@url [1]{\endgroup\@href {#1}{\urlprefix }}%
\providecommand \urlprefix  [0]{URL }%
\providecommand \Eprint [0]{\href }%
\providecommand \doibase [0]{http://dx.doi.org/}%
\providecommand \selectlanguage [0]{\@gobble}%
\providecommand \bibinfo  [0]{\@secondoftwo}%
\providecommand \bibfield  [0]{\@secondoftwo}%
\providecommand \translation [1]{[#1]}%
\providecommand \BibitemOpen [0]{}%
\providecommand \bibitemStop [0]{}%
\providecommand \bibitemNoStop [0]{.\EOS\space}%
\providecommand \EOS [0]{\spacefactor3000\relax}%
\providecommand \BibitemShut  [1]{\csname bibitem#1\endcsname}%
\let\auto@bib@innerbib\@empty
\bibitem [{\citenamefont {Khokhlov}\ \emph {et~al.}(2000)\citenamefont
  {Khokhlov}, \citenamefont {Ivanchik}, \citenamefont {Raines}, \citenamefont
  {Watson},\ and\ \citenamefont {Pipher}}]{Khokhlov-2000}%
  \BibitemOpen
  \bibfield  {author} {\bibinfo {author} {\bibfnamefont {D.~R.}\ \bibnamefont
  {Khokhlov}}, \bibinfo {author} {\bibfnamefont {I.~I.}\ \bibnamefont
  {Ivanchik}}, \bibinfo {author} {\bibfnamefont {S.~N.}\ \bibnamefont
  {Raines}}, \bibinfo {author} {\bibfnamefont {D.~M.}\ \bibnamefont {Watson}},
  \ and\ \bibinfo {author} {\bibfnamefont {J.~L.}\ \bibnamefont {Pipher}},\
  }\href {\doibase http://dx.doi.org/10.1063/1.126489} {\bibfield  {journal}
  {\bibinfo  {journal} {Applied Physics Letters}\ }\textbf {\bibinfo {volume}
  {76}},\ \bibinfo {pages} {2835} (\bibinfo {year} {2000})}\BibitemShut
  {NoStop}%
\bibitem [{\citenamefont {Rogalski}(2003)}]{Rogalski-2003}%
  \BibitemOpen
  \bibfield  {author} {\bibinfo {author} {\bibfnamefont {A.}~\bibnamefont
  {Rogalski}},\ }\href@noop {} {\bibfield  {journal} {\bibinfo  {journal}
  {Progress in Quantum Electronics}\ }\textbf {\bibinfo {volume} {27}},\
  \bibinfo {pages} {59} (\bibinfo {year} {2003})}\BibitemShut {NoStop}%
\bibitem [{\citenamefont {He}\ \emph {et~al.}(2015)\citenamefont {He},
  \citenamefont {Xu}, \citenamefont {Liu}, \citenamefont {Tan}, \citenamefont
  {Shao}, \citenamefont {Liu}, \citenamefont {Xu}, \citenamefont {Jiang},\ and\
  \citenamefont {Jiang}}]{He-2015}%
  \BibitemOpen
  \bibfield  {author} {\bibinfo {author} {\bibfnamefont {J.}~\bibnamefont
  {He}}, \bibinfo {author} {\bibfnamefont {J.}~\bibnamefont {Xu}}, \bibinfo
  {author} {\bibfnamefont {G.}~\bibnamefont {Liu}}, \bibinfo {author}
  {\bibfnamefont {X.}~\bibnamefont {Tan}}, \bibinfo {author} {\bibfnamefont
  {H.}~\bibnamefont {Shao}}, \bibinfo {author} {\bibfnamefont {Z.}~\bibnamefont
  {Liu}}, \bibinfo {author} {\bibfnamefont {J.}~\bibnamefont {Xu}}, \bibinfo
  {author} {\bibfnamefont {J.}~\bibnamefont {Jiang}}, \ and\ \bibinfo {author}
  {\bibfnamefont {H.}~\bibnamefont {Jiang}},\ }\href {\doibase
  10.1039/C5RA08542J} {\bibfield  {journal} {\bibinfo  {journal} {RSC Adv.}\
  }\textbf {\bibinfo {volume} {5}},\ \bibinfo {pages} {59379} (\bibinfo {year}
  {2015})}\BibitemShut {NoStop}%
\bibitem [{\citenamefont {Snyder}\ and\ \citenamefont
  {Toberer}(2008)}]{snyder-2008}%
  \BibitemOpen
  \bibfield  {author} {\bibinfo {author} {\bibfnamefont {G.~J.}\ \bibnamefont
  {Snyder}}\ and\ \bibinfo {author} {\bibfnamefont {E.~S.}\ \bibnamefont
  {Toberer}},\ }\href@noop {} {\bibfield  {journal} {\bibinfo  {journal}
  {Nature materials}\ }\textbf {\bibinfo {volume} {7}},\ \bibinfo {pages} {105}
  (\bibinfo {year} {2008})}\BibitemShut {NoStop}%
\bibitem [{\citenamefont {Dmitriev}\ and\ \citenamefont
  {Zvyagin}(2010)}]{Dmitriev-2010}%
  \BibitemOpen
  \bibfield  {author} {\bibinfo {author} {\bibfnamefont {A.~V.}\ \bibnamefont
  {Dmitriev}}\ and\ \bibinfo {author} {\bibfnamefont {I.~P.}\ \bibnamefont
  {Zvyagin}},\ }\href@noop {} {\bibfield  {journal} {\bibinfo  {journal}
  {Physics-Uspekhi}\ }\textbf {\bibinfo {volume} {53}},\ \bibinfo {pages} {789}
  (\bibinfo {year} {2010})}\BibitemShut {NoStop}%
\bibitem [{\citenamefont {Story}\ \emph {et~al.}(1996)\citenamefont {Story},
  \citenamefont {Sw\"uste}, \citenamefont {Eggenkamp}, \citenamefont
  {Swagten},\ and\ \citenamefont {de~Jonge}}]{Story-1996}%
  \BibitemOpen
  \bibfield  {author} {\bibinfo {author} {\bibfnamefont {T.}~\bibnamefont
  {Story}}, \bibinfo {author} {\bibfnamefont {C.~H.~W.}\ \bibnamefont
  {Sw\"uste}}, \bibinfo {author} {\bibfnamefont {P.~J.~T.}\ \bibnamefont
  {Eggenkamp}}, \bibinfo {author} {\bibfnamefont {H.~J.~M.}\ \bibnamefont
  {Swagten}}, \ and\ \bibinfo {author} {\bibfnamefont {W.~J.~M.}\ \bibnamefont
  {de~Jonge}},\ }\href {\doibase 10.1103/PhysRevLett.77.2802} {\bibfield
  {journal} {\bibinfo  {journal} {Phys. Rev. Lett.}\ }\textbf {\bibinfo
  {volume} {77}},\ \bibinfo {pages} {2802} (\bibinfo {year}
  {1996})}\BibitemShut {NoStop}%
\bibitem [{\citenamefont {Nadolny}\ \emph {et~al.}(2002)\citenamefont
  {Nadolny}, \citenamefont {Sadowski}, \citenamefont {Taliashvili},
  \citenamefont {Arciszewska}, \citenamefont {Dobrowolski}, \citenamefont
  {Domukhovski}, \citenamefont {Łusakowska}, \citenamefont {Mycielski},
  \citenamefont {Osinniy}, \citenamefont {Story}, \citenamefont {Świa̧tek},
  \citenamefont {Gała̧zka},\ and\ \citenamefont {Diduszko}}]{Nadolny-2002}%
  \BibitemOpen
  \bibfield  {author} {\bibinfo {author} {\bibfnamefont {A.}~\bibnamefont
  {Nadolny}}, \bibinfo {author} {\bibfnamefont {J.}~\bibnamefont {Sadowski}},
  \bibinfo {author} {\bibfnamefont {B.}~\bibnamefont {Taliashvili}}, \bibinfo
  {author} {\bibfnamefont {M.}~\bibnamefont {Arciszewska}}, \bibinfo {author}
  {\bibfnamefont {W.}~\bibnamefont {Dobrowolski}}, \bibinfo {author}
  {\bibfnamefont {V.}~\bibnamefont {Domukhovski}}, \bibinfo {author}
  {\bibfnamefont {E.}~\bibnamefont {Łusakowska}}, \bibinfo {author}
  {\bibfnamefont {A.}~\bibnamefont {Mycielski}}, \bibinfo {author}
  {\bibfnamefont {V.}~\bibnamefont {Osinniy}}, \bibinfo {author} {\bibfnamefont
  {T.}~\bibnamefont {Story}}, \bibinfo {author} {\bibfnamefont
  {K.}~\bibnamefont {Świa̧tek}}, \bibinfo {author} {\bibfnamefont
  {R.}~\bibnamefont {Gała̧zka}}, \ and\ \bibinfo {author} {\bibfnamefont
  {R.}~\bibnamefont {Diduszko}},\ }\href {\doibase
  http://dx.doi.org/10.1016/S0304-8853(02)00288-3} {\bibfield  {journal}
  {\bibinfo  {journal} {Journal of Magnetism and Magnetic Materials}\ }\textbf
  {\bibinfo {volume} {248}},\ \bibinfo {pages} {134 } (\bibinfo {year}
  {2002})}\BibitemShut {NoStop}%
\bibitem [{\citenamefont {de~Jonge}\ and\ \citenamefont
  {Swagten}(1991)}]{Jonge-1991}%
  \BibitemOpen
  \bibfield  {author} {\bibinfo {author} {\bibfnamefont {W.}~\bibnamefont
  {de~Jonge}}\ and\ \bibinfo {author} {\bibfnamefont {H.}~\bibnamefont
  {Swagten}},\ }\href {\doibase http://dx.doi.org/10.1016/0304-8853(91)90827-W}
  {\bibfield  {journal} {\bibinfo  {journal} {Journal of Magnetism and Magnetic
  Materials}\ }\textbf {\bibinfo {volume} {100}},\ \bibinfo {pages} {322 }
  (\bibinfo {year} {1991})}\BibitemShut {NoStop}%
\bibitem [{\citenamefont {Hsieh}\ \emph {et~al.}(2012)\citenamefont {Hsieh},
  \citenamefont {Lin}, \citenamefont {Liu}, \citenamefont {Duan}, \citenamefont
  {Bansil},\ and\ \citenamefont {Fu}}]{Hsieh-2012}%
  \BibitemOpen
  \bibfield  {author} {\bibinfo {author} {\bibfnamefont {T.~H.}\ \bibnamefont
  {Hsieh}}, \bibinfo {author} {\bibfnamefont {H.}~\bibnamefont {Lin}}, \bibinfo
  {author} {\bibfnamefont {J.}~\bibnamefont {Liu}}, \bibinfo {author}
  {\bibfnamefont {W.}~\bibnamefont {Duan}}, \bibinfo {author} {\bibfnamefont
  {A.}~\bibnamefont {Bansil}}, \ and\ \bibinfo {author} {\bibfnamefont
  {L.}~\bibnamefont {Fu}},\ }\href {\doibase
  http://dx.doi.org/10.1038/ncomms1969} {\bibfield  {journal} {\bibinfo
  {journal} {Nat. Commun.}\ }\textbf {\bibinfo {volume} {3}},\ \bibinfo {pages}
  {982} (\bibinfo {year} {2012})}\BibitemShut {NoStop}%
\bibitem [{\citenamefont {Tanaka}\ \emph {et~al.}(2012)\citenamefont {Tanaka},
  \citenamefont {Ren}, \citenamefont {Sato}, \citenamefont {Nakayama},
  \citenamefont {Souma}, \citenamefont {Takahashi}, \citenamefont {Segawa},\
  and\ \citenamefont {Ando}}]{Tanaka-2012}%
  \BibitemOpen
  \bibfield  {author} {\bibinfo {author} {\bibfnamefont {Y.}~\bibnamefont
  {Tanaka}}, \bibinfo {author} {\bibfnamefont {Z.}~\bibnamefont {Ren}},
  \bibinfo {author} {\bibfnamefont {T.}~\bibnamefont {Sato}}, \bibinfo {author}
  {\bibfnamefont {K.}~\bibnamefont {Nakayama}}, \bibinfo {author}
  {\bibfnamefont {S.}~\bibnamefont {Souma}}, \bibinfo {author} {\bibfnamefont
  {T.}~\bibnamefont {Takahashi}}, \bibinfo {author} {\bibfnamefont
  {K.}~\bibnamefont {Segawa}}, \ and\ \bibinfo {author} {\bibfnamefont
  {Y.}~\bibnamefont {Ando}},\ }\href {\doibase
  http://dx.doi.org/10.1038/nphys2442} {\bibfield  {journal} {\bibinfo
  {journal} {Nat. Phys.}\ }\textbf {\bibinfo {volume} {8}},\ \bibinfo {pages}
  {800 } (\bibinfo {year} {2012})}\BibitemShut {NoStop}%
\bibitem [{\citenamefont {Yang}\ \emph {et~al.}(2014)\citenamefont {Yang},
  \citenamefont {Liu}, \citenamefont {Fu}, \citenamefont {Duan},\ and\
  \citenamefont {Liu}}]{Yang-2014}%
  \BibitemOpen
  \bibfield  {author} {\bibinfo {author} {\bibfnamefont {G.}~\bibnamefont
  {Yang}}, \bibinfo {author} {\bibfnamefont {J.}~\bibnamefont {Liu}}, \bibinfo
  {author} {\bibfnamefont {L.}~\bibnamefont {Fu}}, \bibinfo {author}
  {\bibfnamefont {W.}~\bibnamefont {Duan}}, \ and\ \bibinfo {author}
  {\bibfnamefont {C.}~\bibnamefont {Liu}},\ }\href {\doibase
  10.1103/PhysRevB.89.085312} {\bibfield  {journal} {\bibinfo  {journal} {Phys.
  Rev. B}\ }\textbf {\bibinfo {volume} {89}},\ \bibinfo {pages} {085312}
  (\bibinfo {year} {2014})}\BibitemShut {NoStop}%
\bibitem [{\citenamefont {Bernick}\ and\ \citenamefont
  {Kleinman}(1970)}]{Bernick-1970}%
  \BibitemOpen
  \bibfield  {author} {\bibinfo {author} {\bibfnamefont {R.}~\bibnamefont
  {Bernick}}\ and\ \bibinfo {author} {\bibfnamefont {L.}~\bibnamefont
  {Kleinman}},\ }\href {\doibase
  http://dx.doi.org/10.1016/0038-1098(70)90305-4} {\bibfield  {journal}
  {\bibinfo  {journal} {Solid State Communications}\ }\textbf {\bibinfo
  {volume} {8}},\ \bibinfo {pages} {569 } (\bibinfo {year} {1970})}\BibitemShut
  {NoStop}%
\bibitem [{\citenamefont {Dimmock}\ \emph {et~al.}(1966)\citenamefont
  {Dimmock}, \citenamefont {Melngailis},\ and\ \citenamefont
  {Strauss}}]{Dimmock-1966}%
  \BibitemOpen
  \bibfield  {author} {\bibinfo {author} {\bibfnamefont {J.~O.}\ \bibnamefont
  {Dimmock}}, \bibinfo {author} {\bibfnamefont {I.}~\bibnamefont {Melngailis}},
  \ and\ \bibinfo {author} {\bibfnamefont {A.~J.}\ \bibnamefont {Strauss}},\
  }\href {\doibase 10.1103/PhysRevLett.16.1193} {\bibfield  {journal} {\bibinfo
   {journal} {Phys. Rev. Lett.}\ }\textbf {\bibinfo {volume} {16}},\ \bibinfo
  {pages} {1193} (\bibinfo {year} {1966})}\BibitemShut {NoStop}%
\bibitem [{\citenamefont {Ferreira}\ \emph {et~al.}(1999)\citenamefont
  {Ferreira}, \citenamefont {Abramof}, \citenamefont {Motisuke}, \citenamefont
  {Rappl}, \citenamefont {Closs}, \citenamefont {Ueta}, \citenamefont
  {Boschetti},\ and\ \citenamefont {Bandeira}}]{Ferreira-1999}%
  \BibitemOpen
  \bibfield  {author} {\bibinfo {author} {\bibfnamefont {S.~O.}\ \bibnamefont
  {Ferreira}}, \bibinfo {author} {\bibfnamefont {E.}~\bibnamefont {Abramof}},
  \bibinfo {author} {\bibfnamefont {P.}~\bibnamefont {Motisuke}}, \bibinfo
  {author} {\bibfnamefont {P.~H.~O.}\ \bibnamefont {Rappl}}, \bibinfo {author}
  {\bibfnamefont {H.}~\bibnamefont {Closs}}, \bibinfo {author} {\bibfnamefont
  {A.~Y.}\ \bibnamefont {Ueta}}, \bibinfo {author} {\bibfnamefont
  {C.}~\bibnamefont {Boschetti}}, \ and\ \bibinfo {author} {\bibfnamefont
  {I.~N.}\ \bibnamefont {Bandeira}},\ }\href@noop {} {\bibfield  {journal}
  {\bibinfo  {journal} {{Brazilian Journal of Physics}}\ }\textbf {\bibinfo
  {volume} {29}},\ \bibinfo {pages} {771 } (\bibinfo {year}
  {1999})}\BibitemShut {NoStop}%
\bibitem [{\citenamefont {Misra}\ and\ \citenamefont
  {Tripathi}(1989)}]{gstpbteandalloy}%
  \BibitemOpen
  \bibfield  {author} {\bibinfo {author} {\bibfnamefont {C.~M.}\ \bibnamefont
  {Misra}}\ and\ \bibinfo {author} {\bibfnamefont {G.~S.}\ \bibnamefont
  {Tripathi}},\ }\href {\doibase 10.1103/PhysRevB.40.11168} {\bibfield
  {journal} {\bibinfo  {journal} {Phys. Rev. B}\ }\textbf {\bibinfo {volume}
  {40}},\ \bibinfo {pages} {11168} (\bibinfo {year} {1989})}\BibitemShut
  {NoStop}%
\bibitem [{\citenamefont {Hota}\ and\ \citenamefont
  {Tripathi}(1991)}]{gstalloy}%
  \BibitemOpen
  \bibfield  {author} {\bibinfo {author} {\bibfnamefont {R.~L.}\ \bibnamefont
  {Hota}}\ and\ \bibinfo {author} {\bibfnamefont {G.~S.}\ \bibnamefont
  {Tripathi}},\ }\href {http://stacks.iop.org/0953-8984/3/i=33/a=009}
  {\bibfield  {journal} {\bibinfo  {journal} {Journal of Physics: Condensed
  Matter}\ }\textbf {\bibinfo {volume} {3}},\ \bibinfo {pages} {6299} (\bibinfo
  {year} {1991})}\BibitemShut {NoStop}%
\bibitem [{\citenamefont {Safaei}\ \emph {et~al.}(2013)\citenamefont {Safaei},
  \citenamefont {Kacman},\ and\ \citenamefont {Buczko}}]{Safaei-2013}%
  \BibitemOpen
  \bibfield  {author} {\bibinfo {author} {\bibfnamefont {S.}~\bibnamefont
  {Safaei}}, \bibinfo {author} {\bibfnamefont {P.}~\bibnamefont {Kacman}}, \
  and\ \bibinfo {author} {\bibfnamefont {R.}~\bibnamefont {Buczko}},\ }\href
  {\doibase 10.1103/PhysRevB.88.045305} {\bibfield  {journal} {\bibinfo
  {journal} {Phys. Rev. B}\ }\textbf {\bibinfo {volume} {88}},\ \bibinfo
  {pages} {045305} (\bibinfo {year} {2013})}\BibitemShut {NoStop}%
\bibitem [{\citenamefont {Rauch}\ \emph {et~al.}(2013)\citenamefont {Rauch},
  \citenamefont {Flieger}, \citenamefont {Henk},\ and\ \citenamefont
  {Mertig}}]{Rauch-2013}%
  \BibitemOpen
  \bibfield  {author} {\bibinfo {author} {\bibfnamefont {T.}~\bibnamefont
  {Rauch}}, \bibinfo {author} {\bibfnamefont {M.}~\bibnamefont {Flieger}},
  \bibinfo {author} {\bibfnamefont {J.}~\bibnamefont {Henk}}, \ and\ \bibinfo
  {author} {\bibfnamefont {I.}~\bibnamefont {Mertig}},\ }\href {\doibase
  10.1103/PhysRevB.88.245120} {\bibfield  {journal} {\bibinfo  {journal} {Phys.
  Rev. B}\ }\textbf {\bibinfo {volume} {88}},\ \bibinfo {pages} {245120}
  (\bibinfo {year} {2013})}\BibitemShut {NoStop}%
\bibitem [{\citenamefont {Gao}\ and\ \citenamefont {Daw}(2008)}]{Gao-2008}%
  \BibitemOpen
  \bibfield  {author} {\bibinfo {author} {\bibfnamefont {X.}~\bibnamefont
  {Gao}}\ and\ \bibinfo {author} {\bibfnamefont {M.~S.}\ \bibnamefont {Daw}},\
  }\href {\doibase 10.1103/PhysRevB.77.033103} {\bibfield  {journal} {\bibinfo
  {journal} {Phys. Rev. B}\ }\textbf {\bibinfo {volume} {77}},\ \bibinfo
  {pages} {033103} (\bibinfo {year} {2008})}\BibitemShut {NoStop}%
\bibitem [{\citenamefont {Zunger}\ \emph {et~al.}(1990)\citenamefont {Zunger},
  \citenamefont {Wei}, \citenamefont {Ferreira},\ and\ \citenamefont
  {Bernard}}]{Zunger-1990}%
  \BibitemOpen
  \bibfield  {author} {\bibinfo {author} {\bibfnamefont {A.}~\bibnamefont
  {Zunger}}, \bibinfo {author} {\bibfnamefont {S.-H.}\ \bibnamefont {Wei}},
  \bibinfo {author} {\bibfnamefont {L.~G.}\ \bibnamefont {Ferreira}}, \ and\
  \bibinfo {author} {\bibfnamefont {J.~E.}\ \bibnamefont {Bernard}},\ }\href
  {\doibase 10.1103/PhysRevLett.65.353} {\bibfield  {journal} {\bibinfo
  {journal} {Phys. Rev. Lett.}\ }\textbf {\bibinfo {volume} {65}},\ \bibinfo
  {pages} {353} (\bibinfo {year} {1990})}\BibitemShut {NoStop}%
\bibitem [{\citenamefont {Wei}\ \emph {et~al.}(1990)\citenamefont {Wei},
  \citenamefont {Ferreira}, \citenamefont {Bernard},\ and\ \citenamefont
  {Zunger}}]{Wei-1990}%
  \BibitemOpen
  \bibfield  {author} {\bibinfo {author} {\bibfnamefont {S.-H.}\ \bibnamefont
  {Wei}}, \bibinfo {author} {\bibfnamefont {L.~G.}\ \bibnamefont {Ferreira}},
  \bibinfo {author} {\bibfnamefont {J.~E.}\ \bibnamefont {Bernard}}, \ and\
  \bibinfo {author} {\bibfnamefont {A.}~\bibnamefont {Zunger}},\ }\href
  {\doibase 10.1103/PhysRevB.42.9622} {\bibfield  {journal} {\bibinfo
  {journal} {Phys. Rev. B}\ }\textbf {\bibinfo {volume} {42}},\ \bibinfo
  {pages} {9622} (\bibinfo {year} {1990})}\BibitemShut {NoStop}%
\bibitem [{\citenamefont {Temmerman}\ \emph {et~al.}(1978)\citenamefont
  {Temmerman}, \citenamefont {Gyorffy},\ and\ \citenamefont
  {Stocks}}]{Temmerman-1978}%
  \BibitemOpen
  \bibfield  {author} {\bibinfo {author} {\bibfnamefont {W.~M.}\ \bibnamefont
  {Temmerman}}, \bibinfo {author} {\bibfnamefont {B.~L.}\ \bibnamefont
  {Gyorffy}}, \ and\ \bibinfo {author} {\bibfnamefont {G.~M.}\ \bibnamefont
  {Stocks}},\ }\href {http://stacks.iop.org/0305-4608/8/i=12/a=008} {\bibfield
  {journal} {\bibinfo  {journal} {Journal of Physics F: Metal Physics}\
  }\textbf {\bibinfo {volume} {8}},\ \bibinfo {pages} {2461} (\bibinfo {year}
  {1978})}\BibitemShut {NoStop}%
\bibitem [{\citenamefont {Faulkner}\ and\ \citenamefont
  {Stocks}(1980)}]{Faulkner-1980}%
  \BibitemOpen
  \bibfield  {author} {\bibinfo {author} {\bibfnamefont {J.~S.}\ \bibnamefont
  {Faulkner}}\ and\ \bibinfo {author} {\bibfnamefont {G.~M.}\ \bibnamefont
  {Stocks}},\ }\href {\doibase 10.1103/PhysRevB.21.3222} {\bibfield  {journal}
  {\bibinfo  {journal} {Phys. Rev. B}\ }\textbf {\bibinfo {volume} {21}},\
  \bibinfo {pages} {3222} (\bibinfo {year} {1980})}\BibitemShut {NoStop}%
\bibitem [{\citenamefont {Ebert}\ \emph {et~al.}(2011)\citenamefont {Ebert},
  \citenamefont {K\"odderitzsch},\ and\ \citenamefont {Min\'{a}r}}]{sprkkr1}%
  \BibitemOpen
  \bibfield  {author} {\bibinfo {author} {\bibfnamefont {H.}~\bibnamefont
  {Ebert}}, \bibinfo {author} {\bibfnamefont {D.}~\bibnamefont
  {K\"odderitzsch}}, \ and\ \bibinfo {author} {\bibfnamefont {J.}~\bibnamefont
  {Min\'{a}r}},\ }\href {http://stacks.iop.org/0034-4885/74/i=9/a=096501}
  {\bibfield  {journal} {\bibinfo  {journal} {Reports on Progress in Physics}\
  }\textbf {\bibinfo {volume} {74}},\ \bibinfo {pages} {096501} (\bibinfo
  {year} {2011})}\BibitemShut {NoStop}%
\bibitem [{\citenamefont {Hohenberg}\ and\ \citenamefont
  {Kohn}(1964)}]{Hohenberg-1964}%
  \BibitemOpen
  \bibfield  {author} {\bibinfo {author} {\bibfnamefont {P.}~\bibnamefont
  {Hohenberg}}\ and\ \bibinfo {author} {\bibfnamefont {W.}~\bibnamefont
  {Kohn}},\ }\href {\doibase 10.1103/PhysRev.136.B864} {\bibfield  {journal}
  {\bibinfo  {journal} {Phys. Rev.}\ }\textbf {\bibinfo {volume} {136}},\
  \bibinfo {pages} {B864} (\bibinfo {year} {1964})}\BibitemShut {NoStop}%
\bibitem [{\citenamefont {Kohn}\ and\ \citenamefont {Sham}(1965)}]{Kohn-1965}%
  \BibitemOpen
  \bibfield  {author} {\bibinfo {author} {\bibfnamefont {W.}~\bibnamefont
  {Kohn}}\ and\ \bibinfo {author} {\bibfnamefont {L.~J.}\ \bibnamefont
  {Sham}},\ }\href {\doibase 10.1103/PhysRev.140.A1133} {\bibfield  {journal}
  {\bibinfo  {journal} {Phys. Rev.}\ }\textbf {\bibinfo {volume} {140}},\
  \bibinfo {pages} {A1133} (\bibinfo {year} {1965})}\BibitemShut {NoStop}%
\bibitem [{\citenamefont {Ernst}(2007)}]{Ernst2007}%
  \BibitemOpen
  \bibfield  {author} {\bibinfo {author} {\bibfnamefont {A.}~\bibnamefont
  {Ernst}},\ }\emph {\bibinfo {title} {Multiple-scattering theory: new
  developments and applications}},\ \href@noop {} {\bibinfo {type}
  {Habilitation thesis}},\ \bibinfo  {school} {Martin Luther University
  Halle-Wittenberg} (\bibinfo {year} {2007})\BibitemShut {NoStop}%
\bibitem [{\citenamefont {L{\"u}ders}\ \emph {et~al.}(2001)\citenamefont
  {L{\"u}ders}, \citenamefont {Ernst}, \citenamefont {Temmerman}, \citenamefont
  {Szotek},\ and\ \citenamefont {Durham}}]{Luders-2001}%
  \BibitemOpen
  \bibfield  {author} {\bibinfo {author} {\bibfnamefont {M.}~\bibnamefont
  {L{\"u}ders}}, \bibinfo {author} {\bibfnamefont {A.}~\bibnamefont {Ernst}},
  \bibinfo {author} {\bibfnamefont {W.}~\bibnamefont {Temmerman}}, \bibinfo
  {author} {\bibfnamefont {Z.}~\bibnamefont {Szotek}}, \ and\ \bibinfo {author}
  {\bibfnamefont {P.}~\bibnamefont {Durham}},\ }\href@noop {} {\bibfield
  {journal} {\bibinfo  {journal} {Journal of Physics: Condensed Matter}\
  }\textbf {\bibinfo {volume} {13}},\ \bibinfo {pages} {8587} (\bibinfo {year}
  {2001})}\BibitemShut {NoStop}%
\bibitem [{\citenamefont {Koelling}\ and\ \citenamefont
  {Harmon}(1977)}]{Koelling-1977}%
  \BibitemOpen
  \bibfield  {author} {\bibinfo {author} {\bibfnamefont {D.~D.}\ \bibnamefont
  {Koelling}}\ and\ \bibinfo {author} {\bibfnamefont {B.~N.}\ \bibnamefont
  {Harmon}},\ }\href@noop {} {\bibfield  {journal} {\bibinfo  {journal}
  {Journal of Physics C: Solid State Physics}\ }\textbf {\bibinfo {volume}
  {10}},\ \bibinfo {pages} {3107} (\bibinfo {year} {1977})}\BibitemShut
  {NoStop}%
\bibitem [{\citenamefont {Takeda}(1978)}]{Takeda-1978}%
  \BibitemOpen
  \bibfield  {author} {\bibinfo {author} {\bibfnamefont {T.}~\bibnamefont
  {Takeda}},\ }\href@noop {} {\bibfield  {journal} {\bibinfo  {journal}
  {Zeitschrift f\"ur Physik B Condensed Matter}\ }\textbf {\bibinfo {volume}
  {32}},\ \bibinfo {pages} {43} (\bibinfo {year} {1978})}\BibitemShut {NoStop}%
\bibitem [{\citenamefont {Marques}\ \emph {et~al.}(2012)\citenamefont
  {Marques}, \citenamefont {Oliveira},\ and\ \citenamefont
  {Burnus}}]{marques-2012}%
  \BibitemOpen
  \bibfield  {author} {\bibinfo {author} {\bibfnamefont {M.~A.}\ \bibnamefont
  {Marques}}, \bibinfo {author} {\bibfnamefont {M.~J.}\ \bibnamefont
  {Oliveira}}, \ and\ \bibinfo {author} {\bibfnamefont {T.}~\bibnamefont
  {Burnus}},\ }\href@noop {} {\bibfield  {journal} {\bibinfo  {journal}
  {Computer Physics Communications}\ }\textbf {\bibinfo {volume} {183}},\
  \bibinfo {pages} {2272} (\bibinfo {year} {2012})}\BibitemShut {NoStop}%
\bibitem [{\citenamefont {Perdew}\ \emph {et~al.}(1996)\citenamefont {Perdew},
  \citenamefont {Burke},\ and\ \citenamefont {Ernzerhof}}]{pbe}%
  \BibitemOpen
  \bibfield  {author} {\bibinfo {author} {\bibfnamefont {J.~P.}\ \bibnamefont
  {Perdew}}, \bibinfo {author} {\bibfnamefont {K.}~\bibnamefont {Burke}}, \
  and\ \bibinfo {author} {\bibfnamefont {M.}~\bibnamefont {Ernzerhof}},\ }\href
  {\doibase 10.1103/PhysRevLett.77.3865} {\bibfield  {journal} {\bibinfo
  {journal} {Phys. Rev. Lett.}\ }\textbf {\bibinfo {volume} {77}},\ \bibinfo
  {pages} {3865} (\bibinfo {year} {1996})}\BibitemShut {NoStop}%
\bibitem [{\citenamefont {Soven}(1967)}]{Soven-1967}%
  \BibitemOpen
  \bibfield  {author} {\bibinfo {author} {\bibfnamefont {P.}~\bibnamefont
  {Soven}},\ }\href@noop {} {\bibfield  {journal} {\bibinfo  {journal} {Phys.
  Rev.}\ }\textbf {\bibinfo {volume} {156}},\ \bibinfo {pages} {809} (\bibinfo
  {year} {1967})}\BibitemShut {NoStop}%
\bibitem [{\citenamefont {Velick\'y}\ \emph {et~al.}(1968)\citenamefont
  {Velick\'y}, \citenamefont {Kirkpatrick},\ and\ \citenamefont
  {Ehrenreich}}]{Velicky-1968}%
  \BibitemOpen
  \bibfield  {author} {\bibinfo {author} {\bibfnamefont {B.}~\bibnamefont
  {Velick\'y}}, \bibinfo {author} {\bibfnamefont {S.}~\bibnamefont
  {Kirkpatrick}}, \ and\ \bibinfo {author} {\bibfnamefont {H.}~\bibnamefont
  {Ehrenreich}},\ }\href@noop {} {\bibfield  {journal} {\bibinfo  {journal}
  {Physical Review}\ }\textbf {\bibinfo {volume} {175}},\ \bibinfo {pages}
  {747} (\bibinfo {year} {1968})}\BibitemShut {NoStop}%
\bibitem [{\citenamefont {Gyorffy}(1972)}]{Gyorffy-1972}%
  \BibitemOpen
  \bibfield  {author} {\bibinfo {author} {\bibfnamefont {B.~L.}\ \bibnamefont
  {Gyorffy}},\ }\href@noop {} {\bibfield  {journal} {\bibinfo  {journal}
  {Physical Review B}\ }\textbf {\bibinfo {volume} {5}},\ \bibinfo {pages}
  {2382} (\bibinfo {year} {1972})}\BibitemShut {NoStop}%
\bibitem [{\citenamefont {Geilhufe}\ \emph {et~al.}(2015)\citenamefont
  {Geilhufe}, \citenamefont {Achilles}, \citenamefont {K\"obis}, \citenamefont
  {Arnold}, \citenamefont {Mertig}, \citenamefont {Hergert},\ and\
  \citenamefont {Ernst}}]{Geilhufe-2015}%
  \BibitemOpen
  \bibfield  {author} {\bibinfo {author} {\bibfnamefont {M.}~\bibnamefont
  {Geilhufe}}, \bibinfo {author} {\bibfnamefont {S.}~\bibnamefont {Achilles}},
  \bibinfo {author} {\bibfnamefont {M.~A.}\ \bibnamefont {K\"obis}}, \bibinfo
  {author} {\bibfnamefont {M.}~\bibnamefont {Arnold}}, \bibinfo {author}
  {\bibfnamefont {I.}~\bibnamefont {Mertig}}, \bibinfo {author} {\bibfnamefont
  {W.}~\bibnamefont {Hergert}}, \ and\ \bibinfo {author} {\bibfnamefont
  {A.}~\bibnamefont {Ernst}},\ }\href@noop {} {\bibfield  {journal} {\bibinfo
  {journal} {arXiv:1506.07743, submitted to Journal of Physics: Condensed
  Matter}\ } (\bibinfo {year} {2015})}\BibitemShut {NoStop}%
\bibitem [{\citenamefont {Vegard}(1921)}]{vegard-1921}%
  \BibitemOpen
  \bibfield  {author} {\bibinfo {author} {\bibfnamefont {L.}~\bibnamefont
  {Vegard}},\ }\href {\doibase 10.1007/BF01349680} {\bibfield  {journal}
  {\bibinfo  {journal} {Zeitschrift f\"ur Physik}\ }\textbf {\bibinfo {volume}
  {5}},\ \bibinfo {pages} {17} (\bibinfo {year} {1921})}\BibitemShut {NoStop}%
\bibitem [{\citenamefont {Mariano}\ and\ \citenamefont
  {Chopra}(1967)}]{Mariano-1967}%
  \BibitemOpen
  \bibfield  {author} {\bibinfo {author} {\bibfnamefont {A.~N.}\ \bibnamefont
  {Mariano}}\ and\ \bibinfo {author} {\bibfnamefont {K.~L.}\ \bibnamefont
  {Chopra}},\ }\href {\doibase http://dx.doi.org/10.1063/1.1754812} {\bibfield
  {journal} {\bibinfo  {journal} {Applied Physics Letters}\ }\textbf {\bibinfo
  {volume} {10}},\ \bibinfo {pages} {282} (\bibinfo {year} {1967})}\BibitemShut
  {NoStop}%
\bibitem [{\citenamefont {Dalven}(1969)}]{Dalven-1969}%
  \BibitemOpen
  \bibfield  {author} {\bibinfo {author} {\bibfnamefont {R.}~\bibnamefont
  {Dalven}},\ }\href {\doibase http://dx.doi.org/10.1016/0020-0891(69)90022-0}
  {\bibfield  {journal} {\bibinfo  {journal} {Infrared Physics}\ }\textbf
  {\bibinfo {volume} {9}},\ \bibinfo {pages} {141 } (\bibinfo {year}
  {1969})}\BibitemShut {NoStop}%
\bibitem [{\citenamefont {Tung}\ and\ \citenamefont {Cohen}(1969)}]{Tung-1969}%
  \BibitemOpen
  \bibfield  {author} {\bibinfo {author} {\bibfnamefont {Y.~W.}\ \bibnamefont
  {Tung}}\ and\ \bibinfo {author} {\bibfnamefont {M.~L.}\ \bibnamefont
  {Cohen}},\ }\href {\doibase 10.1103/PhysRev.180.823} {\bibfield  {journal}
  {\bibinfo  {journal} {Phys. Rev.}\ }\textbf {\bibinfo {volume} {180}},\
  \bibinfo {pages} {823} (\bibinfo {year} {1969})}\BibitemShut {NoStop}%
\bibitem [{\citenamefont {Tung}\ and\ \citenamefont {Cohen}(1970)}]{Tung-1970}%
  \BibitemOpen
  \bibfield  {author} {\bibinfo {author} {\bibfnamefont {Y.~W.}\ \bibnamefont
  {Tung}}\ and\ \bibinfo {author} {\bibfnamefont {M.~L.}\ \bibnamefont
  {Cohen}},\ }\href {\doibase 10.1103/PhysRevB.2.1216} {\bibfield  {journal}
  {\bibinfo  {journal} {Phys. Rev. B}\ }\textbf {\bibinfo {volume} {2}},\
  \bibinfo {pages} {1216} (\bibinfo {year} {1970})}\BibitemShut {NoStop}%
\bibitem [{\citenamefont {Rabii}(1969)}]{Rabii-1969}%
  \BibitemOpen
  \bibfield  {author} {\bibinfo {author} {\bibfnamefont {S.}~\bibnamefont
  {Rabii}},\ }\href {\doibase 10.1103/PhysRev.182.821} {\bibfield  {journal}
  {\bibinfo  {journal} {Phys. Rev.}\ }\textbf {\bibinfo {volume} {182}},\
  \bibinfo {pages} {821} (\bibinfo {year} {1969})}\BibitemShut {NoStop}%
\bibitem [{\citenamefont {Foley}\ and\ \citenamefont
  {Langenberg}(1977)}]{Foley-1977}%
  \BibitemOpen
  \bibfield  {author} {\bibinfo {author} {\bibfnamefont {G.~M.~T.}\
  \bibnamefont {Foley}}\ and\ \bibinfo {author} {\bibfnamefont {D.~N.}\
  \bibnamefont {Langenberg}},\ }\href {\doibase 10.1103/PhysRevB.15.4850}
  {\bibfield  {journal} {\bibinfo  {journal} {Phys. Rev. B}\ }\textbf {\bibinfo
  {volume} {15}},\ \bibinfo {pages} {4850} (\bibinfo {year}
  {1977})}\BibitemShut {NoStop}%
\bibitem [{\citenamefont {Tsu}\ \emph {et~al.}(1968)\citenamefont {Tsu},
  \citenamefont {Howard},\ and\ \citenamefont {Esaki}}]{Tsu-1968}%
  \BibitemOpen
  \bibfield  {author} {\bibinfo {author} {\bibfnamefont {R.}~\bibnamefont
  {Tsu}}, \bibinfo {author} {\bibfnamefont {W.~E.}\ \bibnamefont {Howard}}, \
  and\ \bibinfo {author} {\bibfnamefont {L.}~\bibnamefont {Esaki}},\ }\href
  {\doibase 10.1103/PhysRev.172.779} {\bibfield  {journal} {\bibinfo  {journal}
  {Phys. Rev.}\ }\textbf {\bibinfo {volume} {172}},\ \bibinfo {pages} {779}
  (\bibinfo {year} {1968})}\BibitemShut {NoStop}%
\bibitem [{\citenamefont {Scanlon}(1959)}]{Scanlon-1959}%
  \BibitemOpen
  \bibfield  {author} {\bibinfo {author} {\bibfnamefont {W.}~\bibnamefont
  {Scanlon}},\ }\href {\doibase http://dx.doi.org/10.1016/0022-3697(59)90379-8}
  {\bibfield  {journal} {\bibinfo  {journal} {Journal of Physics and Chemistry
  of Solids}\ }\textbf {\bibinfo {volume} {8}},\ \bibinfo {pages} {423 }
  (\bibinfo {year} {1959})}\BibitemShut {NoStop}%
\bibitem [{\citenamefont {Weinberger}(1990)}]{weinberger-1990}%
  \BibitemOpen
  \bibfield  {author} {\bibinfo {author} {\bibfnamefont {P.}~\bibnamefont
  {Weinberger}},\ }\href@noop {} {\emph {\bibinfo {title} {Electron scattering
  theory for ordered and disordered matter}}}\ (\bibinfo  {publisher}
  {Clarendon Press Oxford},\ \bibinfo {year} {1990})\BibitemShut {NoStop}%
\bibitem [{\citenamefont {Huhne}\ \emph {et~al.}(1998)\citenamefont {Huhne},
  \citenamefont {Zecha}, \citenamefont {Ebert}, \citenamefont {Dederichs},\
  and\ \citenamefont {Zeller}}]{Huhne-1998}%
  \BibitemOpen
  \bibfield  {author} {\bibinfo {author} {\bibfnamefont {T.}~\bibnamefont
  {Huhne}}, \bibinfo {author} {\bibfnamefont {C.}~\bibnamefont {Zecha}},
  \bibinfo {author} {\bibfnamefont {H.}~\bibnamefont {Ebert}}, \bibinfo
  {author} {\bibfnamefont {P.~H.}\ \bibnamefont {Dederichs}}, \ and\ \bibinfo
  {author} {\bibfnamefont {R.}~\bibnamefont {Zeller}},\ }\href {\doibase
  10.1103/PhysRevB.58.10236} {\bibfield  {journal} {\bibinfo  {journal} {Phys.
  Rev. B}\ }\textbf {\bibinfo {volume} {58}},\ \bibinfo {pages} {10236}
  (\bibinfo {year} {1998})}\BibitemShut {NoStop}%
\bibitem [{\citenamefont {Yusheng}\ and\ \citenamefont
  {Grassie}(1985{\natexlab{a}})}]{Yusheng-1985A}%
  \BibitemOpen
  \bibfield  {author} {\bibinfo {author} {\bibfnamefont {H.}~\bibnamefont
  {Yusheng}}\ and\ \bibinfo {author} {\bibfnamefont {A.~D.~C.}\ \bibnamefont
  {Grassie}},\ }\href {http://stacks.iop.org/0305-4608/15/i=2/a=009} {\bibfield
   {journal} {\bibinfo  {journal} {Journal of Physics F: Metal Physics}\
  }\textbf {\bibinfo {volume} {15}},\ \bibinfo {pages} {317} (\bibinfo {year}
  {1985}{\natexlab{a}})}\BibitemShut {NoStop}%
\bibitem [{\citenamefont {Yusheng}\ and\ \citenamefont
  {Grassie}(1985{\natexlab{b}})}]{Yusheng-1985B}%
  \BibitemOpen
  \bibfield  {author} {\bibinfo {author} {\bibfnamefont {H.}~\bibnamefont
  {Yusheng}}\ and\ \bibinfo {author} {\bibfnamefont {A.~D.~C.}\ \bibnamefont
  {Grassie}},\ }\href {http://stacks.iop.org/0305-4608/15/i=2/a=010} {\bibfield
   {journal} {\bibinfo  {journal} {Journal of Physics F: Metal Physics}\
  }\textbf {\bibinfo {volume} {15}},\ \bibinfo {pages} {337} (\bibinfo {year}
  {1985}{\natexlab{b}})}\BibitemShut {NoStop}%
\bibitem [{\citenamefont {Yusheng}\ and\ \citenamefont
  {Graissie}(1985)}]{Yusheng-1985C}%
  \BibitemOpen
  \bibfield  {author} {\bibinfo {author} {\bibfnamefont {H.}~\bibnamefont
  {Yusheng}}\ and\ \bibinfo {author} {\bibfnamefont {A.~D.~C.}\ \bibnamefont
  {Graissie}},\ }\href {http://stacks.iop.org/0305-4608/15/i=2/a=011}
  {\bibfield  {journal} {\bibinfo  {journal} {Journal of Physics F: Metal
  Physics}\ }\textbf {\bibinfo {volume} {15}},\ \bibinfo {pages} {363}
  (\bibinfo {year} {1985})}\BibitemShut {NoStop}%
\bibitem [{\citenamefont {{Wolfram Research, Inc.}}(2014)}]{math}%
  \BibitemOpen
  \bibfield  {author} {\bibinfo {author} {\bibnamefont {{Wolfram Research,
  Inc.}}},\ }\href {https://www.wolfram.com} {\enquote {\bibinfo {title}
  {Mathematica 10.0},}\ } (\bibinfo {year} {2014})\BibitemShut {NoStop}%
\bibitem [{\citenamefont {Ishida}\ \emph {et~al.}(2010)\citenamefont {Ishida},
  \citenamefont {Tsuchiya}, \citenamefont {Yamada}, \citenamefont {Cao},
  \citenamefont {Takaoka}, \citenamefont {Rahim}, \citenamefont {Felder},\ and\
  \citenamefont {Zogg}}]{Ishida-2010}%
  \BibitemOpen
  \bibfield  {author} {\bibinfo {author} {\bibfnamefont {A.}~\bibnamefont
  {Ishida}}, \bibinfo {author} {\bibfnamefont {T.}~\bibnamefont {Tsuchiya}},
  \bibinfo {author} {\bibfnamefont {T.}~\bibnamefont {Yamada}}, \bibinfo
  {author} {\bibfnamefont {D.}~\bibnamefont {Cao}}, \bibinfo {author}
  {\bibfnamefont {S.}~\bibnamefont {Takaoka}}, \bibinfo {author} {\bibfnamefont
  {M.}~\bibnamefont {Rahim}}, \bibinfo {author} {\bibfnamefont
  {F.}~\bibnamefont {Felder}}, \ and\ \bibinfo {author} {\bibfnamefont
  {H.}~\bibnamefont {Zogg}},\ }\href {10.1063/1.3446819} {\bibfield  {journal}
  {\bibinfo  {journal} {Journal of Applied Physics}\ }\textbf {\bibinfo
  {volume} {107}},\ \bibinfo {eid} {123708} (\bibinfo {year}
  {2010})}\BibitemShut {NoStop}%
\bibitem [{\citenamefont {Dixon}\ and\ \citenamefont {Bis}(1968)}]{Dixon-1968}%
  \BibitemOpen
  \bibfield  {author} {\bibinfo {author} {\bibfnamefont {J.~R.}\ \bibnamefont
  {Dixon}}\ and\ \bibinfo {author} {\bibfnamefont {R.~F.}\ \bibnamefont
  {Bis}},\ }\href {\doibase 10.1103/PhysRev.176.942} {\bibfield  {journal}
  {\bibinfo  {journal} {Phys. Rev.}\ }\textbf {\bibinfo {volume} {176}},\
  \bibinfo {pages} {942} (\bibinfo {year} {1968})}\BibitemShut {NoStop}%
\bibitem [{\citenamefont {Yang}(2012)}]{Yang-2012}%
  \BibitemOpen
  \bibfield  {author} {\bibinfo {author} {\bibfnamefont {Y.~L.}\ \bibnamefont
  {Yang}},\ }\href {\doibase 10.1179/1743284712Y.0000000080} {\bibfield
  {journal} {\bibinfo  {journal} {Materials Science and Technology}\ }\textbf
  {\bibinfo {volume} {28}},\ \bibinfo {pages} {1308} (\bibinfo {year}
  {2012})}\BibitemShut {NoStop}%
\bibitem [{\citenamefont {Ovsyannikov}\ \emph {et~al.}(2009)\citenamefont
  {Ovsyannikov}, \citenamefont {Shchennikov}, \citenamefont {Manakov},
  \citenamefont {Likhacheva}, \citenamefont {Ponosov}, \citenamefont
  {Mogilenskikh}, \citenamefont {Vokhmyanin}, \citenamefont {Ancharov},\ and\
  \citenamefont {Skipetrov}}]{Ovsyannikov-2009}%
  \BibitemOpen
  \bibfield  {author} {\bibinfo {author} {\bibfnamefont {S.~V.}\ \bibnamefont
  {Ovsyannikov}}, \bibinfo {author} {\bibfnamefont {V.~V.}\ \bibnamefont
  {Shchennikov}}, \bibinfo {author} {\bibfnamefont {A.~Y.}\ \bibnamefont
  {Manakov}}, \bibinfo {author} {\bibfnamefont {A.~Y.}\ \bibnamefont
  {Likhacheva}}, \bibinfo {author} {\bibfnamefont {Y.~S.}\ \bibnamefont
  {Ponosov}}, \bibinfo {author} {\bibfnamefont {V.~E.}\ \bibnamefont
  {Mogilenskikh}}, \bibinfo {author} {\bibfnamefont {A.~P.}\ \bibnamefont
  {Vokhmyanin}}, \bibinfo {author} {\bibfnamefont {A.~I.}\ \bibnamefont
  {Ancharov}}, \ and\ \bibinfo {author} {\bibfnamefont {E.~P.}\ \bibnamefont
  {Skipetrov}},\ }\href@noop {} {\bibfield  {journal} {\bibinfo  {journal}
  {{Phys. Status Solidi B}}\ }\textbf {\bibinfo {volume} {246}},\ \bibinfo
  {pages} {615 } (\bibinfo {year} {2009})}\BibitemShut {NoStop}%
\bibitem [{\citenamefont {Barone}\ \emph {et~al.}(2013)\citenamefont {Barone},
  \citenamefont {Rauch}, \citenamefont {Di~Sante}, \citenamefont {Henk},
  \citenamefont {Mertig},\ and\ \citenamefont {Picozzi}}]{Barone-2013}%
  \BibitemOpen
  \bibfield  {author} {\bibinfo {author} {\bibfnamefont {P.}~\bibnamefont
  {Barone}}, \bibinfo {author} {\bibfnamefont {T.}~\bibnamefont {Rauch}},
  \bibinfo {author} {\bibfnamefont {D.}~\bibnamefont {Di~Sante}}, \bibinfo
  {author} {\bibfnamefont {J.}~\bibnamefont {Henk}}, \bibinfo {author}
  {\bibfnamefont {I.}~\bibnamefont {Mertig}}, \ and\ \bibinfo {author}
  {\bibfnamefont {S.}~\bibnamefont {Picozzi}},\ }\href {\doibase
  10.1103/PhysRevB.88.045207} {\bibfield  {journal} {\bibinfo  {journal} {Phys.
  Rev. B}\ }\textbf {\bibinfo {volume} {88}},\ \bibinfo {pages} {045207}
  (\bibinfo {year} {2013})}\BibitemShut {NoStop}%
\bibitem [{\citenamefont {Wexler}\ and\ \citenamefont
  {Morgan}(1972)}]{Wexler-1972}%
  \BibitemOpen
  \bibfield  {author} {\bibinfo {author} {\bibfnamefont {G.}~\bibnamefont
  {Wexler}}\ and\ \bibinfo {author} {\bibfnamefont {G.~J.}\ \bibnamefont
  {Morgan}},\ }\href {\doibase 10.1002/pssb.2220490245} {\bibfield  {journal}
  {\bibinfo  {journal} {Physica Status Solidi (b)}\ }\textbf {\bibinfo {volume}
  {49}},\ \bibinfo {pages} {1096} (\bibinfo {year} {1972})}\BibitemShut
  {NoStop}%
\bibitem [{\citenamefont {Radzy\'{n}ski}\ and\ \citenamefont
  {{\L}usakowski}(2009)}]{Radzynski-2009}%
  \BibitemOpen
  \bibfield  {author} {\bibinfo {author} {\bibfnamefont {T.}~\bibnamefont
  {Radzy\'{n}ski}}\ and\ \bibinfo {author} {\bibfnamefont {A.}~\bibnamefont
  {{\L}usakowski}},\ }\href {\doibase
  http://przyrbwn.icm.edu.pl/APP/ABSTR/116/a116-5-62.html} {\bibfield
  {journal} {\bibinfo  {journal} {Acta Phys. Pol. A}\ }\textbf {\bibinfo
  {volume} {116}},\ \bibinfo {pages} {0954} (\bibinfo {year}
  {2009})}\BibitemShut {NoStop}%
\bibitem [{\citenamefont {Qian}\ \emph {et~al.}(2015)\citenamefont {Qian},
  \citenamefont {Fu},\ and\ \citenamefont {Li}}]{Qian-2015}%
  \BibitemOpen
  \bibfield  {author} {\bibinfo {author} {\bibfnamefont {X.}~\bibnamefont
  {Qian}}, \bibinfo {author} {\bibfnamefont {L.}~\bibnamefont {Fu}}, \ and\
  \bibinfo {author} {\bibfnamefont {J.}~\bibnamefont {Li}},\ }\href {\doibase
  10.1007/s12274-014-0578-9} {\bibfield  {journal} {\bibinfo  {journal} {Nano
  Research}\ }\textbf {\bibinfo {volume} {8}},\ \bibinfo {pages} {967}
  (\bibinfo {year} {2015})}\BibitemShut {NoStop}%
\end{thebibliography}

%

\end{document}